\documentstyle[12pt,aasms4]{article}
\tighten

\def\lsim{\lower.5ex\hbox{$\; \buildrel < \over \sim \;$}}
\def\gsim{\lower.5ex\hbox{$\; \buildrel > \over \sim \;$}}
\def\lax    {\ifmmode{_<\atop^{\sim}}\else{${_<\atop^{\sim}}$}\fi}
\def\gax    {\ifmmode{_>\atop^{\sim}}\else{${_>\atop^{\sim}}$}\fi}
\def\etal{{\it et al.\/} }
\def\gtorder{\mathrel{\raise.3ex\hbox{$>$}\mkern-14mu
             \lower0.6ex\hbox{$\sim$}}}
\def\ltorder{\mathrel{\raise.3ex\hbox{$<$}\mkern-14mu
             \lower0.6ex\hbox{$\sim$}}}

\def\pmb#1{\setbox0=\hbox{#1}%
  \kern-0.015em\copy0\kern-\wd0
  \kern0.03em\copy0\kern-\wd0
  \kern-0.015em\raise0.0433em\box0 }

\begin{document}

\title{ Correlations between kHz QPO and Low Frequency Features Attributed
to Radial Oscillations and Diffusive Propagation in the Viscous
 Boundary Layer Around a Neutron Star}

\author{Lev Titarchuk}
\affil{NASA/ Goddard Space
Flight Center, Greenbelt MD 20771, and George Mason University/CSI, USA;
titarchuk@lheavx.gsfc.nasa.gov}

\author{Vladimir Osherovich}
\affil{NASA/Goddard Space Flight Center/RITSS, Greenbelt MD 20771 USA;
vladimir@urap.gsfc.nasa.gov}

\vskip 0.5 truecm


\begin{abstract}

We present a dimensional analysis of two characteristic time scales in
the boundary layer where the disk adjusts to the rotating neutron star (NS).
The boundary layer is treated as a transition region between the
NS surface  and the first Keplerian orbit.
The radial transport of the angular momentum in this layer is controlled
 by a viscous force defined by the Reynolds number, which in turn is
related to the mass accretion rate. We show that the observed low-
Lorentzian frequency is associated with radial oscillations in the
boundary layer, where the observed break frequency is determined by
the characteristic diffusion time of the inward motion of the matter in
the accretion flow. Predictions of our model regarding relations between
those two frequencies and frequencies of kHz QPO's compare favorably
with  recent  observations for the source 4U 1728-34.
This Letter contains a theoretical classification of kHz QPO's
in NS binaries and the related low frequency features.
Thus, results concerning the relationship of the low-Lorentzian frequency
of viscous  oscillations and the break frequency are presented in the
framework of our model of kHz QPO's viewed as Keplerian
oscillations  in a rotating frame of reference.

\end{abstract}

\keywords{accretion, accretion disks---diffusion---stars:individual
(4U 1728-34, Sco X-1)---stars:neutron---X-ray:star---waves}

\section{Introduction}

The discovery of kilohertz quasiperiodic oscillations (QPO's) in the low
mass X-ray neutron star (NS) binaries
(Strohmayer \etal 1996;  Van der Klis \etal 1996 and  Zhang \etal 1996)
has stimulated both theoretical and observational studies of these sources.
In the upper part of the spectrum (400- 1200 Hz) for most of these
sources, two frequencies $\nu_k$ and $\nu_h$ have been seen.
Initially, the fact that for some sources, the peak separation frequency
$\Delta \nu=\nu_h-\nu_k$ does not change much led to the beat frequency
interpretation (Strohmayer \etal 1996;  Van der Klis  1998) which was presented
as a concept for the first time in the paper by Alpar \& Shaham (1985).
Beat-frequency models, where the peak separation is identified with the
NS spin rate have been challenged by observations: for Sco X-1,
$\Delta\nu$ varies by 40\% (van der Klis \etal 1997 hereafter VK97) and 
for source 4U 1608-52, $\Delta\nu$ varies by 26\% (Mendez \etal 1998). 
Mounting observational
evidence that $\Delta\nu$ is not constant demands a new theoretical
approach. For Sco X-1, in the lower part of the spectrum, VK97
identified two branches (presumably the first and second
harmonics) with frequencies 45 and 90 Hz which slowly increase in
frequency when $\nu_k$ and $\nu_h$ increase. Furthermore, in the spectra
observed by Rossi X-ray Timing Explorer (RXTE) for 4U 1728-34, Ford and
van der Klis  (1998, herein FV98) found  low frequency Lorentzian (LFL)
oscillations with frequencies between 10 and 50 Hz. 
These frequencies as well as break frequency, $\nu_{break}$
 of the power spectrum density (PSD)  for the same
source were shown to be correlated with $\nu_k$ and $\nu_h$.
It is clear that the low and high parts of the PSD of the kHz QPO sources
should be related within the framework of the same theory. Difficulties which
the beat frequency model faces are amplified by the requirement of
relating  the observed low frequency features, described above, with
$\nu_k$ and $\nu_h$.

Recently, a different approach to this problem has been suggested:
kHz QPO's in the NS binaries have been modeled by
Osherovich \& Titarchuk (1999) as Keplerian oscillations in a rotating
frame of reference. In this new model the fundamental frequency is the
Keplerian frequency $\nu_k$ (the lower frequency of two kHz QPO's)
\begin{equation}
 \nu_k={{1}\over{2\pi}}\left({{GM}\over{R^3}}\right)^{1/2},
\end{equation}
where G is the gravitational constant, M is the NS mass,
and R is the radius of the corresponding Keplerian orbit.
The high QPO frequency $\nu_h$ is interpreted as the upper hybrid frequency
of the Keplerian oscillator under the influence of the Coriolis force
\begin{equation}
 \nu_h=[\nu_k^2+(\Omega/\pi)^2]^{1/2},
\end{equation}
where $\Omega$ is the angular rotational frequency of the NS magnetosphere.

For three sources (Sco X-1, 4U 1608-52 and 4U 1702-429), we demonstrated
that the solid body rotation ($\Omega=\Omega_0=const$) is a good first
order approximation. Slow variation of $\Omega$ as a function of $\nu_k$
within the second order approximation is related to the differential
 rotation of the magnetosphere controlled by a frozen-in magnetic
structure. This model allows us to address the relation between
the high and low frequency features in the PSD of the neutron systems.
We interpreted the $\sim 45$ and $90$ Hz oscillations as 1st and 2nd
harmonics of the lower branch of the Keplerian oscillations in the rotating
frame of reference:
\begin{equation}
\nu_L=(\Omega/\pi)(\nu_k/\nu_h)\sin\delta,
\end{equation}
where $\delta$ is the angle between ${\bf \Omega}$ and the vector normal
to the plane of the Keplerian oscillations. For Sco X-1, we found that the
angle $\delta=5.5^o$  fits the observations.

In this Letter we include the LFL oscillations  and related break frequency 
phenomenon in our classification. 
We attribute LFL oscillations to radial oscillations in
the  viscous boundary layer surrounding a neutron star. According to the
model of Shakura \& Sunyaev (1973, hereafter SS73), 
the innermost part of the Keplerian
disk adjusts itself to the rotating central object (i.e. neutron star).
The recent modelling by Titarchuk, Lapidus \& Muslimov (1998, hereafter TLM)
led to the determination of the characteristic thickness of the viscous
boundary layer $L$. In the following section, we present the extension
of this work to relate the frequency of the viscous oscillations $\nu_v$
and $\nu_{break}$ with $\nu_k$. Comparison with the observations is carried
out for 4U 1728-34. The last section of this Letter contains our
theoretical classification of kHz QPO's and related low
frequency phenomena.
\section{Radial Oscillations and Diffusion in the Viscous Boundary Layer}
We define the boundary layer as a transition region confined between the
NS surface  and the first Keplerian orbit.
The radial motion in the disk is controlled by the friction and the angular
momentum exchange between adjacent layers resulting in the loss of the
initial angular momentum by an accreting matter. The corresponding radial
transport of the angular momentum in a disk is described by the equation
(e.g. SS73):
\begin{equation}
 \dot M {d\over {dR}}(\omega R^2) =
2\pi {d\over {dR}} (W_{r\varphi}R^2),
\end{equation}
where $\dot{M}$ is the accretion rate, and $ W_{r\varphi}$ is
the component of a viscous stress tensor which is related to the gradient
of the rotational frequency $\omega$, namely
\begin{equation}
 W_{r\varphi}=-2\eta HR{{d\omega}\over{dR}},
\end{equation}
where $H$ is a half-thickness of a disk, and $\eta$ is the turbulent
viscosity. The nondimensional parameter which is essential for
equation (4) is the Reynolds number for the accretion flow
\begin{equation}
 \gamma={{\dot M}\over{4\pi\eta H}}={{3R v_r}\over {{\it v}_t{\it l}_t}},
\end{equation} 
which is the inverse   $\alpha-$parameter in the SS73-model;
$v_r$ is a characteristic velocity, $v_t$ and $l_t$ are a turbulent velocity
 and related turbulent scale respectively.

  Equations $\rm \omega=\omega_0~{\rm~at}~R=R_0$ (NS radius) and
  ${\rm \omega=\omega_K~at~R=R_{out}}$ (radius where the boundary layer 
adjusts to the Keplerian motion),
and $\rm {{d\omega}\over{dr}}={{d\omega_k}\over{dr}}~
at~\rm R=R_{out}$ were assumed by TLM as boundary conditions.
Thus the profile $\omega(R)$ and the outer radius of the viscous
boundary layer $R_{out}$ are uniquely determined by these boundary
conditions.

Presenting $\omega(R)$ in terms of dimensionless
variables: namely
 angular velocity $\theta=\omega/\omega_0$, radius $ r=R/R_0$
($ R_0=x_0R_s$, $ R_s=2GM/c^2$ is the Schwarzschild radius), and mass
$ m=M/M_{\odot}$, we express Keplerian angular
velocity as 
\begin{equation}
\theta_K={{6}/(a_K r^{3/2}}),
\end{equation}
where  $a_K=m(x_0/3)^{3/2}(\nu_0/363~{\rm Hz})$
%
and the NS rotational frequency $\nu_0$ has a particular value 
for each star.
The particular coefficient, 6, presented in formula (7) is obtained
for the frequency of nearly coherent (burst) oscillations for 4U 1728-34,
i.e. for $\nu_0=363$ Hz.
The solution of equations (4-5 ) satisfying the above
 boundary conditions  is
\begin{equation}
 \theta(r)=D_1 r^{-\gamma} + (1-D_1) r^{-2},
\end{equation}
where  $D_1=(\theta_{out}-r_{out}^{-2})/(r_{out}^{-\gamma}
-r_{out}^{-2})$  and $\theta_{out}=\theta_K(r_{out})$.
%
Equation $\theta^{\prime}(r_{out})=\theta_K^{\prime}(r_{out})$ determines
$r_{out}$:
\begin{equation}
{3\over2} \theta_{out}=D_1 \gamma r_{out}^{-\gamma}+2(1-D_1)r_{out}^{-2}.
\end{equation}

The solution of equations (4-5) subject to the
inner sub-Keplerian boundary condition  has a
regime corresponding to the super-Keplerian rotation (TLM).
For such a regime  matter piles up in the
vertical direction thus disturbing the hydrostatic equilibrium. 
The vertical component of the gravitational
force prevents this matter from further accumulation in a vertical
direction and drives relaxation oscillations. The radiation drag
force, which is proportional to the vertical velocity,
determines the characteristic decay time of the vertical oscillations 
(TLM).
The characteristic time $t_r$,  over which the matter moves inward through
this region, bounded between the innermost disk and relaxation 
oscillations zone is
\begin{equation}
t_r\sim {L\over v_r},
\end{equation}
where $L=R_{out}-R_0$ is the characteristic thickness of this region.
Even though the specific mechanism providing the modulation of the
observed X-ray flux over this timescale needs to be
understood, this timescale apparently ``controls'' the supply of
accreting matter into the innermost region of the accretion disk.
Any local perturbation in the transition region would
propagate diffusively outward over a timescale
\begin{equation}
t_{diff}\sim \left({L\over {l_{fp}}}\right)^2 {{l_{fp}}\over v_r},
\end{equation}
where $l_{fp}$ is the mean free path of a particle.

Note, that the $\gamma-$parameter is proportional to the accretion rate
(see Eq. 6), and therefore $v_r\propto \gamma$. Using this
relationship, we can exclude $v_r$ from the above equations and
get the relations for the corresponding inverse timescales (frequencies).
For the frequency of viscous oscillations
\begin{equation}
\nu_{v}\propto {\gamma\over {r_{out}-r_0}},
\end{equation}
and for the break frequency, related to the diffusion
\begin{equation}
\nu_{break}\propto{\gamma\over {(r_{out}-r_0)^2}}.
\end{equation}

In the following section, we compare the predictions of this model with
the observations and also establish the theoretical relation between
$\nu_v$ and $\nu_{break}$.

\section{Comparisons with Observations}

The results of FV98 for the low frequency
Lorentzian in the X-ray binary 4U 1728-34 are presented in Figure 1 and 
for the break frequency $\nu_{break}$ in Figure 2. 
In Figure 1, crosses represent the frequencies
(with the appropriate error bars) observed during four days.
Data collected on February 16 (open circles) are situated
apart from the rest of observations and they are not included
in the empirical power law fit which is suggested by FV98.
In the  work discussed above, the authors plotted the observed low
frequencies versus high-frequency QPO which for all days, except 
February 16 was
$\nu_k$ and apparently for February 16 it was $\nu_h$. Our theoretical
curve for $\nu_v$ versus $\nu_k$ is based on equation (12).
The $\chi^2$ dependence
on this parameter  is rather strong: the parabola
$\chi^2=38024-73076\cdot a_k+35732\cdot a_k^2$ has a minimum at
$a_k=1.03$, which determines the best fit.
Using $\Omega/2\pi=340$ Hz in the upper hybrid relation (2), 
we calculate $\nu_k$
for the points observed on February 16 and show that they belong to the set
of frequencies modeled by our theoretical curve for the viscous radial
oscillations (closed circles).

Identification of the observed
$\nu_{break}$ with the inverse diffusion time (formulas 11 and 13) is
illustrated by theoretical curves in Figure 2. It is worth noting that
these two correlations with kHz frequencies are fit by two theoretical
curves using {\it only one parameter $a_k$}. The
$\chi^2-$ dependence on $a_k$  is obtained with inclusion of
all data points for the break and low frequency correlations (75 data points).
 The theoretical dependences of $\nu_k$ and  $r_{out}$ on $\gamma-$parameter
are calculated numerically using equations (7) and (9) and  employed here for
calculations of the theoretical curves in Figures 2 and 3 using equations
(12) and (13).  

We were unable to interpret data for $\nu_{break}$ collected in February 16
(open circles). Neither $\nu_v$ nor $\nu_{break}$ in our theory have a power 
law relation with $\nu_k$. However, the theoretical relation between 
$\nu_{break}$ and $\nu_v$, shown in Figure 3 by a solid curve, is close 
to the straight line (in log-log diagram), suggesting an approximate power law
\begin{equation}
\nu_{break}=0.041\nu_v^{1.61}.
\end{equation}
This relation is derived from the theoretical dependence
for the best fit parameter $a_k=1.03$. 
Observations of FV98 (except February 16) are also presented in the Figure 3.

\centerline {\bf 4.  Discussion and Conclusions}

We present a model for the radial oscillations and diffusion in the
viscous boundary layer  surrounding the neutron star. Our dimensional
analysis has identified the corresponding frequencies $\nu_v$ and
$\nu_{break}$ which are consistent with
the low Lorentzian and break frequencies  for 4U 1728-34. and 
 predicted values for $\nu_{break}$ related to the diffusion 
in the boundary layer are consistent with the break
frequency observed for the same source. Both
oscillations (Keplerian and radial) and diffusion in the viscous boundary
layer are controlled by the same parameter - Reynolds number $\gamma$
which in turn is related to the accretion rate.  
It is shown in TLM that $\nu_k$ is a monotonic function of  $\gamma$. 
Therefore, the observed range of $\nu_k$, (350-900 Hz) corresponds to
the range $1<\gamma<5$ (or $0.2<\alpha<1$). 

The results in this Letter extend the classification of kHz QPO's and the
related low frequency phenomena suggested by Osherovich \& Titarchuk
1999. Figure 4 summarizes the new classification. Solid lines represent our
theoretical curves and open circles observations for Sco X-1
(from VK97). As one can see, formulas (2) and (3),
for the Keplerian oscillator under the influence of the Coriolis force,
reproduce the observations well. Indeed, $\Delta\nu=\nu_h-\nu_k$ is not
constant, as observed (see OT99 for details of comparisons of the data 
with the theory). Effectively, the main viscous frequency $\nu_v$ and the
diffusive $\nu_{break}$ introduce the second oscillator with
 two new branches in the lower part of the spectra.
The unifying characteristic of spectra for both oscillators is the strong
dependence on $\nu_k$. This common dependence on $\nu_k$ can be viewed as a
result of the interaction between Keplerian oscillator and the viscous
oscillator which share the common boundary at the outer edge of the viscous 
transition layer.
 
Our parametric study indicates
that the power law index 1.6 in Eq. (14) should be the same 
for different neutron stars. We expect a similar relation for black holes
but with a distinctly different index. 

The found value of $a_k$  leads ultimately to  independent
constraints in the determination of mass and radius for the neutron star
(Haberl \& Titarchuk 1995).

LT thanks NASA for support under grants NAS-5-32484 and
RXTE Guest Observing Program. The authors acknowledge discussions
with Alex Muslimov, Jean Swank, Lorella Angelini,  Will Zhang, 
Joe Fainberg and fruitful suggestions by  the referee.  
Particularly, we are grateful to Eric Ford, and Michiel van der Klis, 
for the data which enable us to make comparisons with the data
in detail.

\clearpage

\begin{figure}
\caption{ Main Viscous Frequency  (crosses)
versus the frequency of the kHz QPOs for 4U 1728-34
(FV98). Open circles are for the  data
of February 16. Solid circles are for the data of 16th of February
recalculated using the Keplerian frequencies with the rotational frequency
340 Hz (see text).
 The solid line is the  best theoretical fit
for the main viscous frequency calculated using Eqs. (4-5, 9, 12) 
corresponding to the parameter $a_k=1.03$.
\label{Fig.1}}
\end{figure}

\begin{figure}
\caption{ Break Frequency   of a broken power law function of
the power spectrum density (crosses)
versus the frequency of the kHz QPOs for 4U 1728-34
(FV98). Open circles are for the  data
of February 16.
 The solid line is the  theoretical
curve  for break frequencies calculated using Eqs. (4-5, 9, 13) corresponding
to the parameter $a_k=1.03$.
\label{Fig.2}}
\end{figure}

\begin{figure}
\caption{Break frequency versus main viscous frequency
(stars) for  all data points except those of February 16.
Solid line is the theoretical curve which is almost a power law with
index 1.61.
\label{Fig.3}}
\end{figure}

\begin{figure}
\caption{Classification of the QPO frequencies in kHz QPO sources. Open 
circles of QPO data by VK97.
\label{Fig.4}}
\end{figure}

\noindent
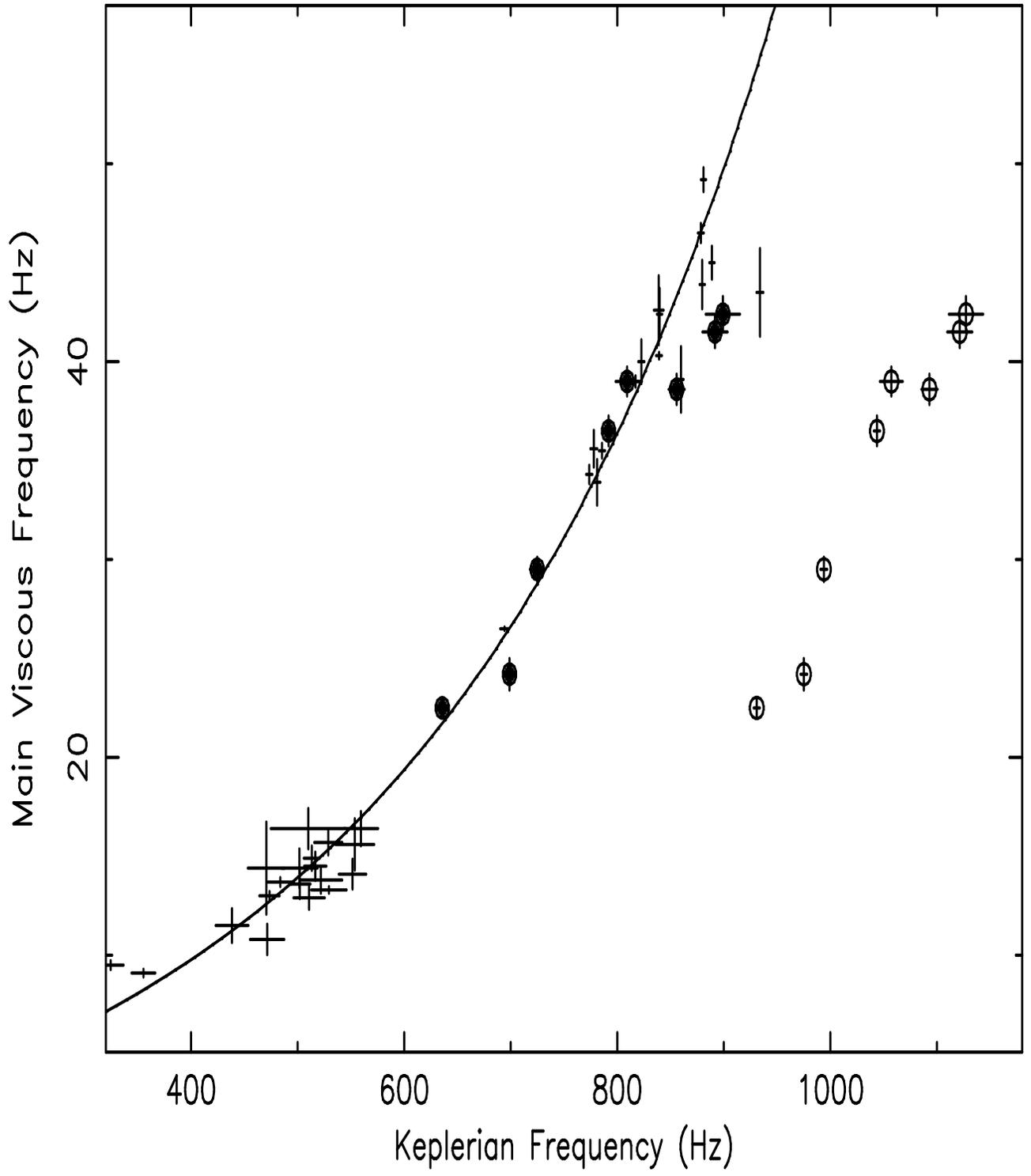
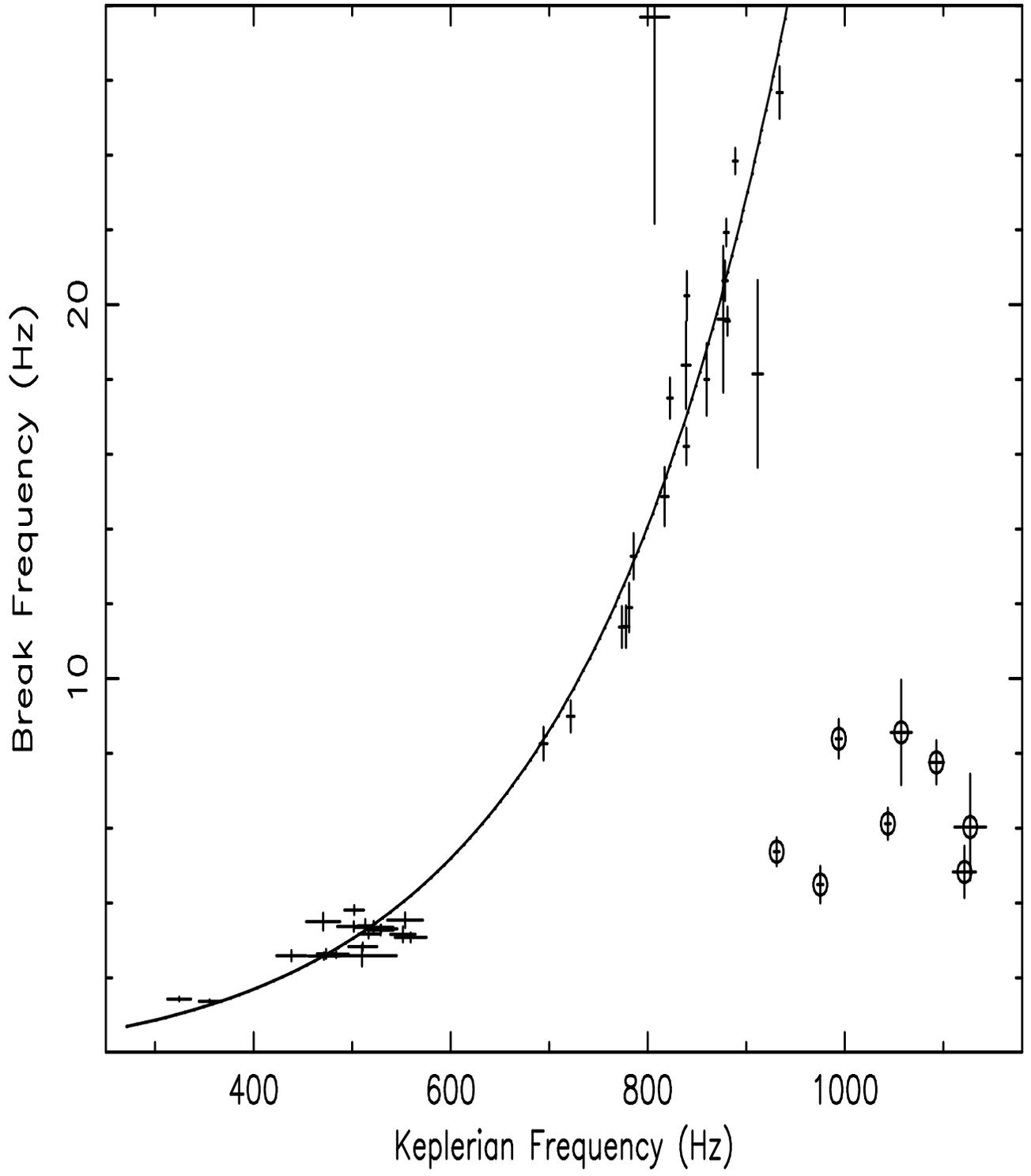
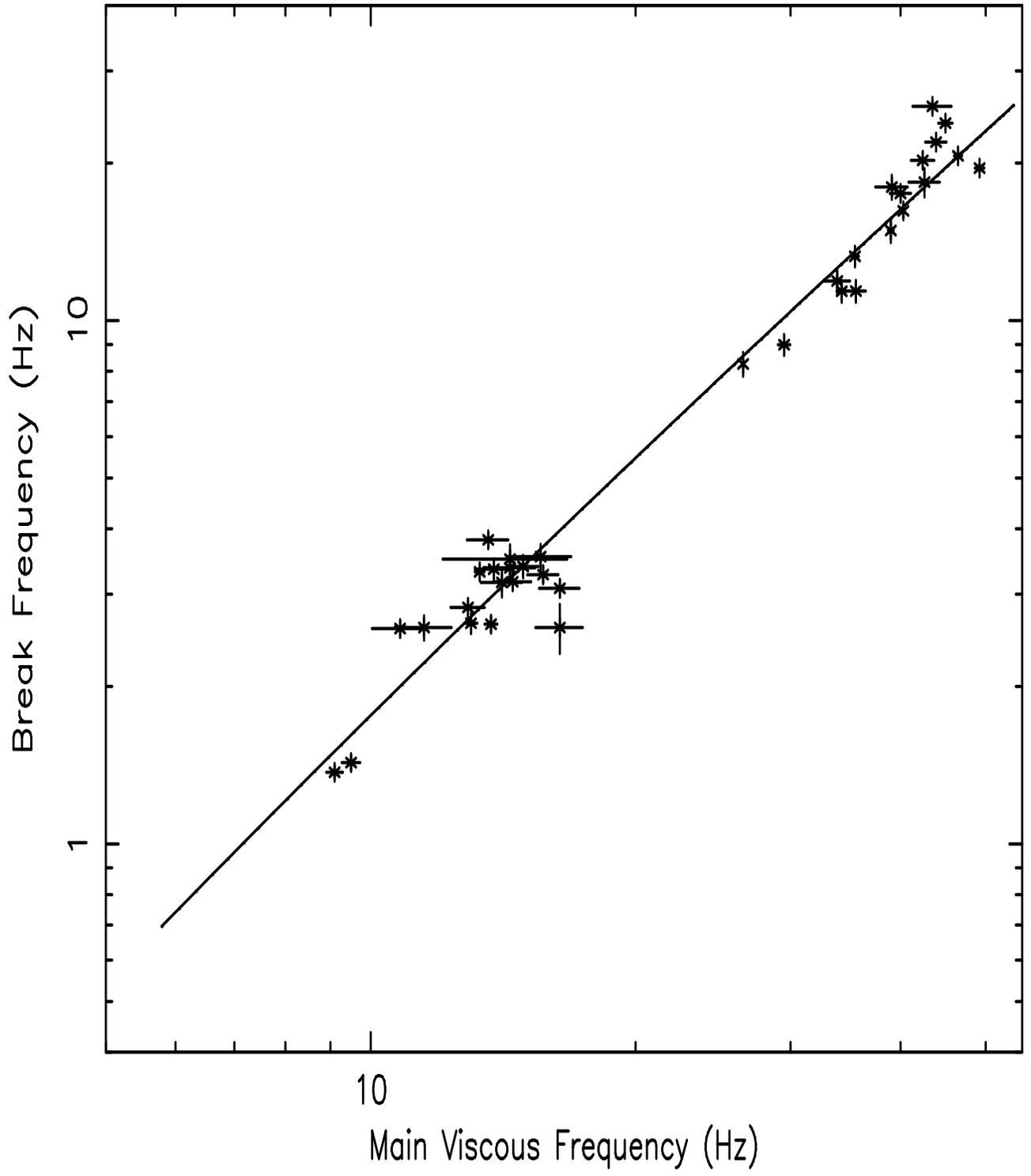
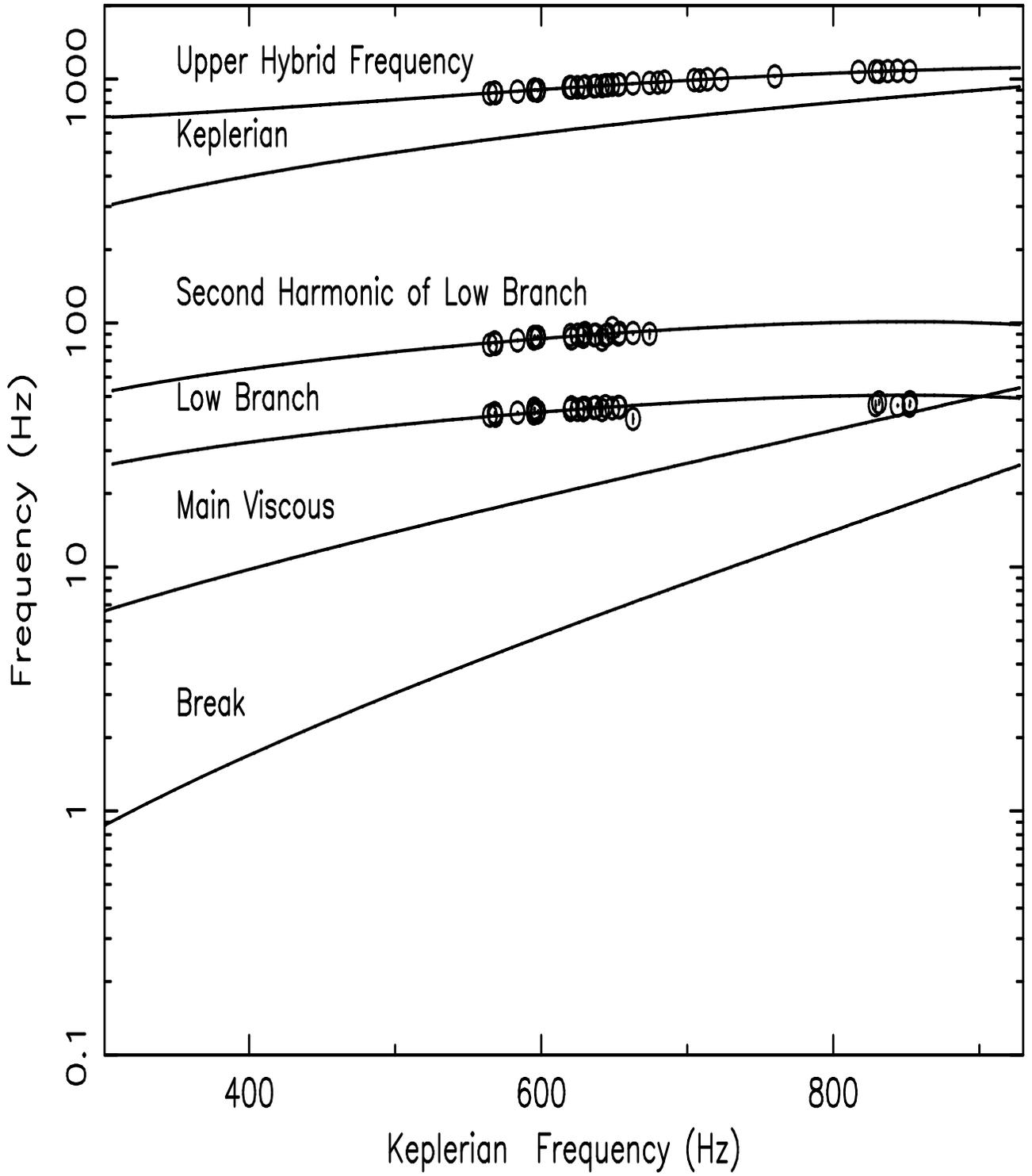
\end{document}